\documentclass[superscriptaddress,aps,prl,twocolumn]{revtex4}
\usepackage{graphicx}
\usepackage{amsmath}
\usepackage{bm}
\usepackage{color}
\begin{document}
\title{Principle of maximum force and holographic principle:\\ two principles or one?}
\author{Yu.\,L.\,Bolotin}
\email{ybolotin@gmail.com}
\author{V.\,A.\,Cherkaskiy}
\email{vcherkaskiy@gmail.com}
\affiliation{A.I.Akhiezer Institute for Theoretical Physics,
National Science Center "Kharkov Institute of Physics and
Technology", Akademicheskaya Str. 1, 61108 Kharkov, Ukraine}
\date{\today}
\begin{abstract}
We show how the maximum force principle can be derived from the holographic principle and vice versa, thus demonstrating equivalence of the two principles.
\end{abstract}
\maketitle
Information is material, i.e. it always needs a material carrier. Information is impossible without matter. Let us consider some physical system of macroscopic dimensions, the Universe for example, different parts of which contain some information. Let our task be to concentrate a part of the information available inside some finite volume during the shortest possible time interval. What principal limitations shall we face when performing this task?

Concentration of information inside certain volume inevitably requires to concentrate its material carriers inside the same volume. The existing physical limitations on concentration of energy lead to direct forbidding of the excessive information concentration process. It is a direct consequence of the fact that the information must have a material carrier. In the most laconic form these two types of the limitations were formulated as the maximum force principle \cite{gibbons,schiller,barrow_gibbons} and the holographic principle \cite{t_hooft,susskind}. The present paper attempts to prove the equivalence of these two principles or at least a deep connection between them.

The Principle of Maximum Force states that the tension or force between two bodies cannot exceed the value
\begin{equation}\label{eq1}
F_{max}=\frac{c^4}{4G}\approx3.25\times10^{43}N.
\end{equation}
This limit does not depend on nature of the forces and is valid for gravitational, electromagnetic, nuclear or any other forces. Absolutely equivalent statement is the existence of the maximum power
\begin{equation}\label{eq2}
P_{max}=\frac{c^5}{4G}\approx9.07\times10^{51}W.
\end{equation}
Maximum force and maximum power are invariants, which follows from invariance of the quantities $c$ and $G$. Its time dependence however is not excluded.

There is one more equivalent limit --- mass change rate in nature is limited by the maximum value
\begin{equation}\label{eq3}
\frac{dm}{dt}\le\frac{c^3}{4G}\approx1.0009\times10^{35}kg/sec.
\end{equation}
Existence of the maximum force is generally speaking a principle (postulate), but it is quite clear how it comes to being. Let us consider the attractive gravitational force between two masses $M_1$ and $M_2$, separated by the distance $D$:
\begin{align}
\nonumber & F= G\frac{M_1M_2}{D^2} = \left(\frac{GM_1}{c^2D}\right) \left(\frac{GM_2}{c^2D}\right)\frac{c^4}G,\\
\nonumber & M_1M_2\le\frac14(M_1+M_2)^2,\\
\label{eq4} & F\le\left[\frac{(M_1+M_2)G}{c^2D}\right]^2\frac{c^4}{4G},\\
\nonumber & (M_1+M_2)G\le c^2D,\\
\nonumber & F\le\frac{c^4}{4G}.
\end{align}
The maximum force value is attained at the moment of creation of the event horizon for the black hole of mass $M_1+M_2$, i.e. when the separation $D$ between the masses $M_1$ and $M_2$ becomes equal to the corresponding black hole size. The surfaces which realize the maximum force (the maximum momentum flow) or the maximum power (the maximum energy flow) are horizons. A horizon appears at any attempt to surpass the force limit. The horizon prohibits the possibility to surpass the limit.

The holographic principle is based on the connection between entropy and information. Amount of information $I$ connected with some matter and its position (in any microscopic theory) is measured in terms of the entropy
\begin{equation}\label{eq5}
\Delta S=-\Delta I.
\end{equation}
Variation of entropy with displacement of matter lads to the so-called entropic force \cite{verlinde,easson_frampton_smoot}. Its origin therefore lies in the universal tendency of any microscopic theory to maximize the entropy. There are no fundamental fields associated with the entropic forces, and dynamic equations are immediately expressed in terms of the entropy variation.

The simplest formulation of the holographic principle contains the two statements:
\begin{enumerate}
\item All information contained in some region of space can be ''recorded'' (represented) on the boundary of this region, which is called the holographic screen;
\item A theory on the boundary of the considered region of space must contain not more than one degree of freedom per Planck area, or in other words total number of degrees of freedom $N$ satisfies the inequality
\begin{equation}\label{eq6}
N\le\frac{A}{4l_{Pl}^2}=\frac{Ac^3}{4G\hbar}.
\end{equation}
\end{enumerate}
It means that the information density on the holographic screen is limited by the value $1/l_{Pl}^2\approx10^{69}bit/m^2$.

Principle of the maximum force and the holographic principle are related due to the fact that both of them represent a statement about existence of some maximum value for a certain quantity: the force (or power) for the former and information density for the latter case. The statement about the existence of limiting values can be used as a basis for physical axiomatics. Thus for example, quantum mechanics can be constructed on the assumption about the existence of minimum quantum of action $\hbar$, special relativity --- on the existence of the maximum velocity $c$, and general relativity --- on the existence of the maximum force (\ref{eq1}). In all the above cited examples the actual value of the limiting quantity matters less than the importance of the fact of its very existence.

When comparing the principle of the maximum force and the holographic principle one realizes that both of them have common origin --- the event horizon \cite{culetu}. This connection is reflected in the so-called IR-UV correspondence \cite{schiller_mountain,bolotin_yerokhin_lemets}, which allowed to introduce the notion of the holographic dark energy \cite{cohen_kaplan_nelson}: total energy in a volume of linear dimensions $L$ cannot exceed mass of the black hole of the same size
\[L^3\rho_{DE}\le LM_{Pl}^2,\]
where $\rho_{DE}$ is the dark energy density and $M_{Pl}$ is the Planck mass.

A natural question arises: why Nature uses two dominant principles instead of one? For the first sight, one should determine ''the most important'' of them, in order to be safe of contradictions between them if they appear. Such contradictions already took place and led to serious crises: recall reversibility of mechanics (or relativity) and time arrow (in thermodynamics). The Nature is quite astute to avoid such contradictions. It seems to provide that those two principles are linked to each other in some way, though enigmatic for us.

This connection is a manifestation of the fact that the description of physical processes (of Nature) in terms of the information represents an alternative to the traditional one in terms of the physical laws. Information processing in the physical systems obeys a number of fundamental limitations \cite{lloyd}. It is those limitations which can be considered as an origin to appearance of the limiting values in form of the maximum force principle. Of course one can legally treat the holographic principle as a consequence of the maximum force one.

Let us first attempt to obtain the maximum power value (\ref{eq2}) using the fundamental limitations of the information theory. In the formulas below one should understand the notion $X\approx Y$ as $\log X=\log Y+O(1)$. For that purpose we will need the following fundamental limits:
\begin{enumerate}
\item The Margolus-Levitin theorem \cite{margolus_levitin} states that the total number of elementary logical operations, which a system can perform per unit time, is limited by the average excess of energy over the ground state
\begin{equation}\label{eq7}
N_{ops/sec}\le\frac{2E}{\pi\hbar}.
\end{equation}
\item Total number of bits available to a system is limited by its entropy
\begin{equation}\label{eq9}
N_{bit}\le \frac S{k_B\ln2}
\end{equation}
\item The Landauer principle \cite{landauer}: any computing system regardless of its physical realization in order to process a bit of information needs energy amount limited by
\begin{equation}\label{eq10}
E_{bit}\ge E_{SNL}\equiv k_B T\ln2,
\end{equation}
where $E_{SNL}$ stands for the Shannon-von Neumann-Landauer energy.
In other words, temperature can be considered as the average energy of one bit of information on the holographic screen.
\end{enumerate}
Treating processing of one bit as an elementary logical operation, one finds that
\begin{equation}\label{eq11}
N_{max}\approx N_{ops/sec}\times E_{SNL}\approx \frac E\hbar k_BT.
\end{equation}
As we have seen above, the maximum values are attained exclusively on the horizons. Taking for the horizon temperature the Hawking value
\begin{equation}\label{eq0}
T_H=\frac{\hbar c^3}{8\pi k_B GM}
\end{equation}
and using that $E=Mc^2$, one obtains
\begin{equation}\label{eq12}
N_{max}\approx \frac{c^5}G,
\end{equation}
which corresponds to the postulated value of the maximum power (\ref{eq2}).

We have shown above how to obtain the maximum power value (\ref{eq12}) starting from the holographic principle. Now we make the opposite route: without any information theory limitations, we shall show that the information density on the holographic screen is limited by the quantity \[\frac1{l_{Pl}^2}=\frac{c^3}{G\hbar}=10^{69}bit/m^2.\]

As Schiller \cite{schiller_mountain} shows, a direct consequence of the maximum force principle is the so-called horizon equation
\begin{equation}\label{eq13}
E=\frac{c^2}{8\pi G}aA.
\end{equation}
This equation relates the energy flow through the are $A$ of the spherical horizon with the surface gravity \[a=\frac{c^2}R,\] where $R$ is the horizon radius. The relation (\ref{eq13}) states that the energy flow through the horizon is limited and that the energy is proportional to the area and to the surface gravity. After substitution of $E=N_{bit}k_BT\ln2$ into (\ref{eq13}) one obtains
\begin{equation}\label{eq14}
\frac{N_{bit}}A=\frac1{k_BT}\frac{c^2}{8\pi G}\,a\approx\frac1{l_{Pl}^2}=\frac{c^3}{G\hbar}=10^{69}bit/m^2.
\end{equation}
Like in the derivation of (\ref{eq12}), we took into account that the maximum values are attained only on the horizons, therefore we took the horizon temperature equal to the Hawking one.

The obtained equivalence of the maximum force principle to the holographic one can be presented in a more general form. The horizon equation (\ref{eq13}) for a horizon that is irregularly curved or time-varying becomes \cite{schiller_mountain}
\begin{equation}\label{eq15}
\delta E=\frac{c^2}{8\pi G}a\delta A.
\end{equation}
The energy flow $\delta E$ through the local Rindler horizon of the accelerated observer and the corresponding entropy change satisfy the following balance equation \cite{jacobson} 
\begin{equation}\label{eq16}
\delta S=\frac{\delta E}T,
\end{equation}
where $T$ represents the horizon temperature which we take equal to the Unruh temperature \cite{unruh}, related to the surface gravity by
\begin{equation}\label{eq17}
T=\frac{\hbar c}{2\pi k_B}a.
\end{equation}
It then follows from Eq. (\ref{eq15})-(\ref{eq17}) that
\begin{equation}\label{eq18}
\delta S=\frac{c^3}{4G\hbar}\delta A=k_B\frac{\delta A}{4l_{Pl}^2}.
\end{equation}
The latter expression represents a direct result of the holographic principle (\ref{eq6}).

To conclude, we obtain the maximum force value (\ref{eq2}) starting from the holographic principle and vice versa, i.e. we show that the qualitative characteristic of the holographic screen --- the information density --- can be derived from the maximum force principle. Thus we demonstrate the equivalence between the maximum force principle and the holographic principle, implying that the two actually can be considered as one. We essentially use the Landauer principle in order to establish the material equivalent of the information.

\end{document}